\documentstyle[twoside,fleqn,espcrc2,epsf]{article}
\title{
\thispagestyle{empty}
\vspace{-14mm}
\rightline{\small ITEP-LAT/2001-05}
\rightline{\small 15 October, 2001}
\vspace{2mm}
Theta term instead of the Higgs field in Electroweak theory
\thanks{Presented by M.A.Z. at Lattice 2001, Berlin.}}

\author{M.I.~Polikarpov \address[ITEP]{Institute of Theoretical and
        Experimental Physics,
        B.Cheryemushkinskaya 25, Moscow, 117259 Russia},
        A.I.~Veselov \addressmark[ITEP], M.A.~Zubkov \addressmark[ITEP]
}

\begin{document}
\sloppy
\begin{abstract}
We consider the electroweak theory without
fundamental scalar field.
The topological excitation of the
$SU(2)\times U(1)$ theory (the monopole) plays
the role of the Higgs field,
it carries the $SU(2)\times U(1)$ topological charge due
to the theta--term of the special type.
\end{abstract}

\maketitle

In this paper we suggest a new formulation of Wainberg - Salam model on the
lattice. The nice feature of our approach is that we have no fundamental
Higgs field in the initial Lagrangian (there been a number of attempts to
eliminate the fundamental Higgs field from electroweak Lagrangian
\cite{C,A}). The Higgs field appears as topological monopole excitation;
due to $\theta$--term with $\theta = \pi$ the $U(1)$ monopole acquires the
$SU(2)\times U(1)$ charge and become the scalar electrically charged field
(see also \cite{W,V}). The theory may contain phase in which $U(1)$
monopoles are condensed and they play the role of the Higgs field in the
electroweak theory.

The Villain action
of the $U(1)$ model with the theta-term has the form:

\begin{eqnarray}
S = \sum_{plaq} \beta |d\theta +2\pi M|^2 \nonumber\\
- i/(4\pi) (d \theta + 2\pi M, ^* (d \theta +  2\pi M)).
\end{eqnarray}
Here $\theta$ is $U(1)$ gauge field, $j = ^*d M$ is the monopole worldline,
$^*$ is the
duality transformation, $M$ is the Villain variable. Thus we have
\begin{eqnarray}
S = \sum_{plaq} \beta |d\theta +2\pi M|^2
\nonumber\\
+ i (\theta,j) +i\pi (M,^*M).
\end{eqnarray}

Note that monopole currents $j$ interact with the gauge field $\theta$ via
the Wilson loop, $i(\theta,j)$. Thus monopoles carry the $U(1)$ charge. The
partition function of the theory has the form

\begin{eqnarray}
Z = \int^{\pi}_{-\pi} D\theta
\sum_{j=*dM}
 exp(-\sum_{plaq}
\nonumber\\
\beta |d\theta +2\pi M|^2
 - i (\theta,j) - i\pi (M, ^*M))
\nonumber\\
  = \int^{\infty}_{-\infty} D\theta
 \sum_{\delta j = 0} exp(-\sum_{plaq} \beta |d\theta |^2  \nonumber\\
 - U(j) - i (\theta,j))
\end{eqnarray}

Here $U(j)$ is a nonlocal potential.
Next we can transform the sum over the monopoles into the integral
over the scalar field.
\begin{eqnarray}
Z = \int^{\infty}_{-\infty} D\theta  D\Phi
 exp(-\sum_{plaq} \beta |d\theta |^2 -V(|\Phi|)\nonumber\\
 - \sum_{xy} (\Phi_x^+ e^{i \theta_{xy}}-
 \Phi^+_y)(\Phi_x e^{-i \theta_{xy}}- \Phi_y))
\end{eqnarray}

Here $V(r)$ is a nonlocal potential.

This example illustrates how the scalar electrically charged field appears
from the theory containing monopoles.
Our aim now is to generalize this construction to the nonabelian theory.

First we consider the continuum electroweak theory.
The fermion part
of the action looks like (the fermions are supposed to be massless)
\begin{equation}
S = S^{left}_{leptons}+S^{right}_{leptons}
\end{equation}

\begin{equation}
S^{left}_{leptons} = \int \bar{L}(\partial^{\mu} + iA^{\mu}_{SU(2)}
+ iB^{\mu}_{U(1)})\gamma_{\mu} L d^4x
\nonumber
\end{equation}

\begin{equation}
S^{right}_{leptons} = \int \bar{R}(\partial^{\mu}
+ 2iB^{\mu}_{U(1)})\gamma_{\mu} R d^4x
\end{equation}

The physical variables depend upon parallel transporters
along the closed lines in the appropriate representations.
Thus the additional symmetry takes place:
\begin{eqnarray}
A^i \rightarrow  A^i + (A^j t^j/\sqrt{tr (A^j t^j_{^*v})^2})
\pi (^*v)^i
\nonumber \\
B^i \rightarrow  B^i  -  \pi (^* v)^i
\end{eqnarray}
Here  $v$ is the 3-dimensional hypersurface,
$t^i_{^*v}$ is the unity vector in the direction $^*v$.

\begin{equation}
v^{ijk} = \int_{v} \delta(x-x(a,b,c)) dx^i \Lambda dx^j \Lambda dx^k
\end{equation}
$t^{i}=\epsilon^{ijkl} (dx^k/da)(dx^l/db)(dx^j/dc)$.

This additional symmetry appears since we consider the "compact" continuum
theory in which the singular gauge fields are not forbidden. The compact
continuum nonabelian fields are discussed in ref. \cite{zakh} as the limit
of lattice gauge fields.
The lattice analogue of this symmetry is
\begin{eqnarray} \label{symm}
U \rightarrow -U;~~
\theta \rightarrow \theta + \pi
\end{eqnarray}
for $U \in SU(2), \theta \in U(1)$.

\begin{figure}[htb]
\vspace{-3cm} \hspace{.5cm}
\epsfxsize=6.5cm\epsfysize=7.cm\epsffile{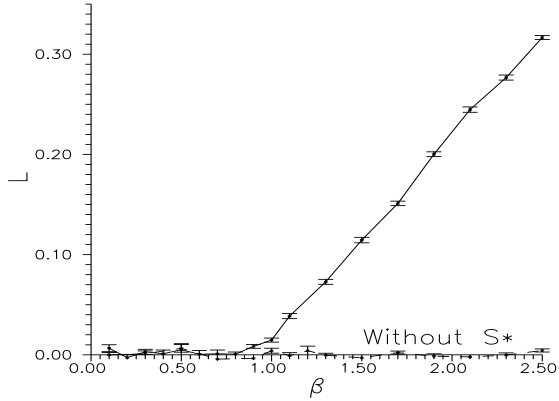}
\vspace{-0.5cm}
\caption{Polyakov line and confinement - deconfinement phase
transition.  Dots "without $S^*$" represent the results for the theory
without theta-term.} \label{fig:largenenough} \end{figure}

This symmetry corresponds to the center $Z_2$ of $SU(2)$ group,
and it can be shown \cite{BVZ} that due to this symmetry the so-called
center monopoles coincide with the ordinary $U(1)$ monopoles in the
continuum. The action for the pure gauge fields should also posses this
symmetry. The above mentioned symmetry prompts us the following definition
of the $SU(2) \times U(1)$ theta - term \begin{eqnarray} Q = 1/4\pi \int
R^*R d^4x \\ R_{ij} = \partial_i B_j -\partial_j B_i + H_{ij} + 2\pi
\epsilon_{ijkl} \Sigma_{kl}, \end{eqnarray} Here $H$ is t'Hooft tensor
\begin{equation} H_{ij} = tr ( G_{ij} n) - tr( n D_i n D_j n),
\end{equation} $G_{ij}=[D_i,A_j]$, $D_j=\partial_j + A_j$, $n = n^a
\sigma^a$.

$\Sigma_{ij}$ is the antisymmetric tensor representing
the 2- dimensional surface $\Sigma$. This tensor is the analogue of the
Villain variable from the lattice formulation of $U(1)$ theory with
theta-term.

The integration over $\Sigma$ and $n$ $(n^a n^a=1)$ is assumed.

\begin{equation}
\Sigma^{ij} = \int_{\Sigma} \delta(x-x(a,b)) dx^i \Lambda dx^j
\end{equation}

We have the model with the partition function
\begin{eqnarray}
Z = \int DA DB Dn D\Sigma exp(-S_{fermions}\nonumber\\
- S_{gauge fields} + i Q)\nonumber
\end{eqnarray}
We found that
\begin{eqnarray}
Z = \int DA DB D j \nonumber\\ exp(-S_{fermions} - S_{gauge fields})
\nonumber\\
P exp ( i \int_j (A_i - B_i)dx_i) e^{-V(j)}
\end{eqnarray}
Here $V(j) $ is nonlocal potential, $j = \delta \Sigma$ is
the monopole worldline.

We can express the sum over the worldlines of the monopoles $j$ as
the integral over the scalar field

\begin{eqnarray}
Z = \int DA DB D\Phi exp (S_{fermions} - S_{gauge fields})\nonumber\\
- \int|(d+iA-iB)\Phi|^2 - V(|\Phi|) \nonumber
\end{eqnarray}
Here $V(|\Phi|)$ is unknown potential.
It means that the low energy approximation of our
theory can be equivalent to the Wainberg - Salam model for the
massless fermions if the potential $V$ is of the Higgs type.

We can express the continuum partition function in the following way:
\begin{eqnarray}
Z = \int DA DB Dg Dj \nonumber\\
exp(-S_{fermions} - S_{gauge fields})
exp (  \int_j A^g_3)
\end{eqnarray}
We use this expression to construct the lattice theory
\begin{eqnarray}
Z = \int DU D\theta D g \sum_{\delta j=0}  exp (-S_f-S_g)\nonumber\\
\Pi_{link \in j}((U^g)^{11}_{link} exp(i\theta_{link}))
\nonumber
\end{eqnarray}
Here
\begin{eqnarray}
S_g = \beta \sum_{plaquettes} ((1-1/2 tr U_p cos \theta_p)\nonumber\\
+(1-cos2\theta_p)\nonumber
\end{eqnarray}
The lattice theory possesses the above considered additional symmetry
(\ref{symm}).

In numerical calculations we use the lattice $10^4-16^4$. At large values of
$\beta$ the confinement is destroyed. It follows from the evaluation of the
Creutz ratios and the Polyakov line.
The confinement - deconfinement phase transition is the of 1-st order.
It follows from the calculation of the correlation:

\begin{eqnarray}
\rho(|x-y|) = <\sum_{x\in plaq}(trU_{plaq} cos\theta_{plaq})
\nonumber\\
\sum_{y\in plaq}(trU_{plaq} cos\theta_{plaq})>
\nonumber\\- <\sum_{x\in plaq}(trU_{plaq} cos\theta_{plaq})>^2\nonumber
\end{eqnarray}

\begin{equation}
\rho(r) \rightarrow Const  *  exp(-M r)
\end{equation}
We found that $M$ is almost independent on $\beta$.
That means that the correlation length does not tend to infinity at the point
of the phase transition. Thus the theory under consideration has
no continuum limit.

Our conclusions are:

1. We consider the $SU(2)\times U(1)$ gauge theory. We found that
the fermion part of Wainberg-Salam model (for massless fermions)
possesses the additional $Z_2$ symmetry. We suppose that this
symmetry plays the important role in the construction of the action
for the gauge field.

2. We construct the $SU(2) \times U(1)$ theta - term of the special type.

3. The resulting theory contains monopoles, which carry the
$SU(2) \times U(1)$ charge. The monopoles are condensed by the construction.
The  low energy approximation of the theory can be equivalent to
the Wainberg - Salam model. The monopole field plays the role of
the Higgs field.

4. We formulate a lattice version of the theory.
It occurs that this version has no continuum limit.
It means that we should redefine the lattice action to obtain
the second order phase transition.

5. We do not discuss how the fermion masses appear in our model.

6. We do not discuss how $M_W$ and $\theta_W$ appear in our model.

\vspace{0.25cm}
The work has been supported by grants
RFBR 01-02-17456, RFBR 00-15-96786, RFBR 99-01-01230, INTAS 00-00111 and
CRDF award RP1-2103.

\end{document}